\documentclass[aps,prl,amsmath,showpacs,twocolumn]{revtex4}

\usepackage[latin1]{inputenc}
\usepackage[T1]{fontenc}
\usepackage[dvips]{graphicx}
\usepackage{color}
\usepackage[UKenglish]{babel}
\usepackage{subfigure}
\usepackage{ulem}

\newcommand{\RANGLE}%
{\mathchoice{\bigr\rangle}{\bigr\rangle}{\rangle}{\rangle}}
\newcommand{\LANGLE}%
{\mathchoice{\bigl\langle}{\bigl\langle}{\langle}{\langle}}

\begin{document}

\title{Field-free orientation of CO molecules by femtosecond two-color laser fields}

\author{S. De$^1$, I. Znakovskaya$^2$, D. Ray$^1$, F. Anis$^1$, Nora G. Johnson$^1$, I. A. Bocharova$^1$, M. Magrakvelidze$^1$, B. D. Esry$^1$, C. L. Cocke$^1$, I. V. Litvinyuk$^1$, M. F. Kling$^{1,2}$}

\affiliation{$^1$J.R. Macdonald Laboratory, Physics Department, Kansas State University, Manhattan, KS 66506, USA, Email: ivl@phys.ksu.edu
$^2$Max-Planck Institute of Quantum Optics, Hans-Kopfermann-Str. 1, 85748 Garching, Germany, Email: matthias.kling@mpq.mpg.de}

\begin{abstract}
We report the first experimental observation of non-adiabatic field-free orientation of a heteronuclear diatomic molecule (CO) induced by an intense two-color (800 and 400~nm) femtosecond laser field. We monitor orientation by measuring fragment ion angular distributions after Coulomb explosion with an 800~nm pulse. The orientation of the molecules is controlled by the relative phase of the two-color field. The results are compared to quantum mechanical rigid rotor calculations. The demonstrated method can be applied to study molecular frame dynamics under field-free conditions in conjunction with a variety of spectroscopy methods, such as high-harmonic generation, electron diffraction and molecular frame photoelectron emission.
\end{abstract}
\maketitle
\begin{figure}
 \centering
 \includegraphics[trim=0 20 0 10,width=0.8\columnwidth]{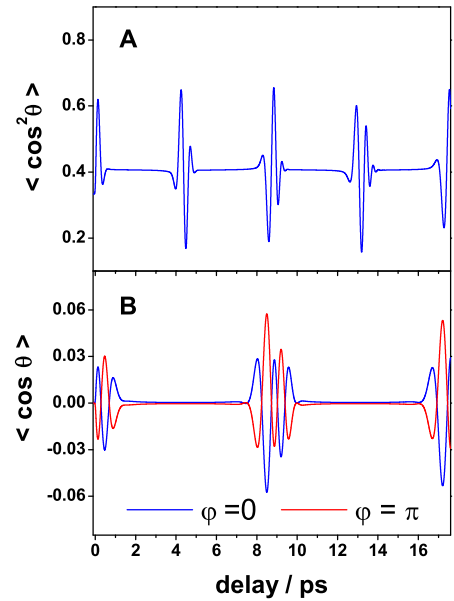}
 \caption{Theoretically obtained time evolution of the alignment parameter <cos$^2~\theta$> and the orientation parameter <cos~$\theta$> for the two-color (800~nm and 400~nm) excitation of CO.}
 \label{fig:orienttheory}
\end{figure}

Aligned molecules have attracted widespread interest for applications such as ultrafast dynamic imaging \cite{Ivanov07}, molecular tomography \cite{Itatani04}, and electron diffraction \cite{Meckel08}. Molecules have been successfully aligned in one and even three dimensions using strong linearly polarized laser fields \cite{Viftrup07}. Aligned, but unoriented, gas samples of heteronuclear diatomic molecules, however, suffer from an averaging effect over the two opposite molecular orientations. To overcome this limitation, it is necessary to develop methods for their orientation. While the orientation of polar molecules is possible in strong DC fields (see e.g. \cite{Friedrich91}) or by the combination of a laser field and a DC field \cite{Friedrich99,Minemoto03,Holmegaard09}, the presence of a strong field may influence the outcome of experiments on oriented molecular targets. It is thus of particular interest to study and implement methods for field-free orientation and its control. In this letter, we report on the non-adiabatic field-free orientation of a heteronuclear diatomic molecule (CO) using a two-color laser field.
\begin{figure}
 \centering
 \includegraphics[trim=0 0 0 0,angle=-90,width=\columnwidth]{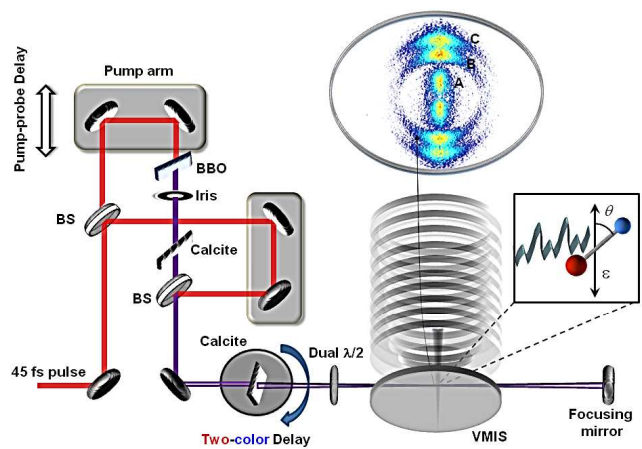}
 \caption{Schematic of the experimental setup used to induce and monitor field-free molecular orientation. A,B and C indicate channels for the formation of C$^{2+}$ ions as described in the text.}
 \label{fig:setup}
\end{figure}
The control of molecular alignment can be achieved both adiabatically \cite{Tanji05} and non-adiabatically \cite{Stapelfeldt03}.
Non-adiabatic alignment can be used to produce aligned samples of molecules under field-free conditions and, hence, is desirable for some applications. In laser-induced orientation, in addition, a head versus tail order of the molecules is established. Only recently, two groups have demonstrated laser-field-free transient molecular orientation \cite{goban08,ghafur09}. Sakai and coworkers have used a combination of a DC electric field and a laser pulse with adiabatic turn-on and non-adiabatic turn-off (switched laser field) to induce dynamic orientation in OCS molecules, which revives at full rotational periods \cite{goban08}. Vrakking and coworkers used a hexapole state-selector to produce NO molecules in a single quantum state and a combination of a DC field and an intense femtosecond laser field to induce orientation \cite{ghafur09} of their samples. Although in both of these cases laser-field-free oriented molecules were obtained at full rotational revivals, the presence of a DC field was crucial to achieving orientation and might limit the application of these techniques in e.g. the imaging of low-energy photoelectrons from oriented molecules. Another "all-optical" route to field-free orientation without the necessivity of a DC field was suggested by Kanai and Sakai \cite{SakaiJCP2001} and has been further explored theoretically, most recently by Tehini and Sugny \cite{tehini08} and by Sakai and coworkers \cite{Muramatsu09}. The approach is based on the nonadiabatic excitation of both odd and even angular momentum states with a femtosecond two-color laser field, enabling net macroscopic orientation \cite{vrakking97}. So far, the technique has not been experimentally studied.
\\
In this letter, we describe experiments where an intense two-color laser field
\begin{equation}
E(t)=E_{\omega}(t) \cos(\omega t) + E_{2 \omega}(t) \cos(2 \omega t +  \varphi)
\end{equation}
with wavelengths 800 and 400~nm corresponding to $\omega$ and 2$\omega$ and the phase $\varphi$ was used to control the field-free orientation of carbon monoxide. The degree of orientation was probed by Coulomb explosion imaging. The setup is displayed in Fig. \ref{fig:setup}. Pulses with 45~fs duration at 800~nm produced from a Ti:Sapphire laser were split into a pump and a probe arm of a Mach-Zehnder interferometer. In the pump arm, the second harmonic of 800~nm was created using a 250~$\mu$m thick Beta Barium Borate (BBO) crystal. To ensure temporal overlap of the two colors, two calcite plates of thickness 600~$\mu$m were used. Together both the calcite crystals compensate the group delay between the two colors caused by the BBO crystal while the second calcite plate also serves to adjust the relative phase $\varphi$ between the two colors of the excitation field. The phase was calibrated by comparison of above-threshold ionization data from Xe to quantitative rescattering theory calculations \cite{micheau09_short}. A dual $\lambda$/2 plate (which rotates the fundamental and the second-harmonic fields by 90$^{\circ}$ and 180$^{\circ}$, respectively) was used to rotate the polarization of all fields to be vertical in the experiment. An iris placed in the pump arm gives control over the intensity of the excitation field. The resulting field-asymmetric two-color excitation pulses (schematically shown for $\varphi$=0 in the inset of Fig. \ref{fig:setup}) were focused onto a supersonic jet of CO molecules (T$_{rot} \approx$ 60 K) inside a velocity-map imaging spectrometer (VMIS) by a spherical mirror (f~=~75~mm) placed at the rear side of the VMIS. The CO molecules were Coulomb exploded at a varying time delay by a single-color (800 nm) laser pulse from the probe arm and the resulting fragment ions were projected by ion optics of the VMIS onto a MCP-phosphor screen assembly. Images were recorded with a CCD camera. Note that a DC field on the ion optics orthogonal to the laser polarization direction was used to image the fragment ions, however, this DC field does not take any role in the orientation of the molecules and could even be made zero for e.g. the imaging of electrons from oriented molecules by zero-field time-of-flight spectroscopy.
\\
A typical image of C$^{2+}$ ions from the Coulomb explosion of CO recorded in our experiment is shown in Fig. \ref{fig:setup}. The image shows three main contributions peaked along the vertical laser polarization axis, which we attribute to different channels for the production of C$^{2+}$ ions, where in addition to C$^{2+}$ a neutral (A), singly charged (B) and doubly charged (C) oxygen is formed. The energy and angular distributions of the fragments can be directly derived from the image after inversion using an iterative inversion procedure \cite{vrakking01}. The angle $\theta$ is defined as the angle between the momentum of C$^{2+}$ ions and the laser polarization direction. For Coulomb explosion channels (B and C in Fig. \ref{fig:setup}), we assume the fragmentation occurs along the molecular axis (axial recoil approximation). Within this approximation $\theta$ reflects the angle between the molecular axis and the laser polarization as depicted in the inset of Fig. \ref{fig:setup}.
\begin{figure}
 \centering
 \includegraphics[trim=0 20 0 10,width=0.8\columnwidth]{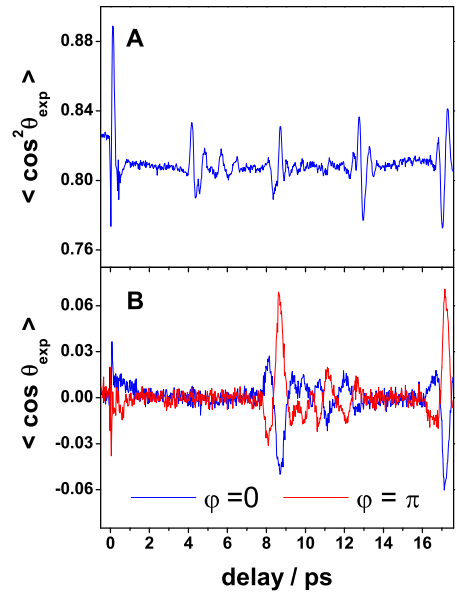}
 \caption{Evolution of the (A) alignment parameter <cos$^2~\theta_{\text{exp}}$> and (B) orientation parameter <cos~$\theta_{\text{exp}}$> with pump probe delay time for two opposite phases $\varphi$ of the two-color pump.}
 \label{fig:alor}
\end{figure}
Fig. \ref{fig:alor}A shows a trace of the experimental parameter <cos$^2~\theta_{\text{exp}}$> versus the two-color pump - Coulomb explosion probe delay time for $\varphi$=0. The parameter <cos$^2~\theta_{\text{exp}}$> was obtained by integration over the Coulomb explosion channel B in Fig. \ref{fig:setup} corresponding to a kinetic energy range of 10 to 16~eV and reflects the alignment of the molecules. The data was taken for peak intensities of 1.3$\cdot$10$^{14}$W/cm$^2$ for both the 800 and 400~nm pump pulses and 2.4$\cdot$10$^{14}$W/cm$^2$ for the 800~nm probe pulse. No significant dependence of the alignment parameter on the phase $\varphi$ between the two colors of the pump pulse was found. The alignment parameter <cos$^2~\theta_{\text{exp}}$> peaks at $T_0$ = 0.125~ps and shows the first two full revivals at $T = T_0 + n \cdot T_{rot}$ = 8.7 and 17.3~ps for n = 1, 2 in good agreement with the expected rotation time $T_{rot} = 1 /(2 B_0 c)$ of 8.64 ps (with B$_0$ = 1.93 cm$^{-1}$) \cite{NIST} for CO and also in good agreement with earlier studies \cite{pinkham05}. The alignment curve also shows half-revivals at 4.3 and 12.9~ps. Before interaction with the pump, i.e. at negative delay times, we already find a high value of <cos$^2~\theta_{\text{exp}}$> of 0.83 (an isotropic angular distribution corresponds to a value of 0.33), which we attribute to the additional alignment caused by the linearly polarized probe pulse \cite{Litvinyuk03}. The alignment by the probe, however, cannot affect an existing up-down asymmetry in the ion image and therefore does not influence the results of the current studies on molecular orientation.
\begin{figure}
 \centering
 \includegraphics[trim=0 20 0 0,width=0.95\columnwidth]{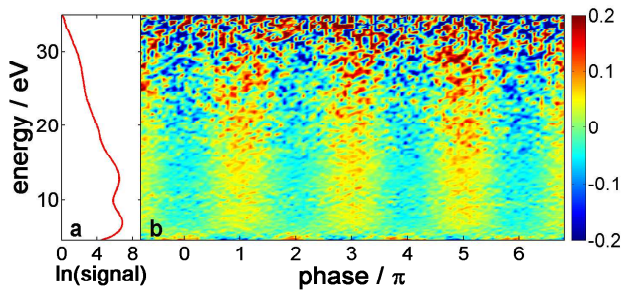}
 \caption{(A) Kinetic energy spectrum of C$^{2+}$ ions obtained by full angular integration for $\varphi$ = 0 at the first full revival (8.7 ps) of the alignment and (B) map of <cos~$\theta_{\text{exp}}$> at this delay as a function of the C$^{2+}$ kinetic energy and the phase $\varphi$ of the two-color laser field.}
 \label{fig:orientmap}
\end{figure}
In addition to the non-adiabatic alignment of CO seen in Fig. \ref{fig:alor}A, we investigated the non-adiabatic orientation with the two-color laser field and the possible control of the orientation direction by the phase $\varphi$. Along these lines, we performed a pump - probe experiment at a fixed delay between the two pulses corresponding to the first full revival of the alignment (at 8.7~ps) and recorded <cos~$\theta_{\text{exp}}$> parameter as a function of the kinetic energy of the resulting C$^{2+}$ fragments and the phase $\varphi$.
\\
Fig. \ref{fig:orientmap}A shows the relevant part of the kinetic energy spectrum of recorded C$^{2+}$ ions (obtained by full angular integration) for the Coulomb explosion channels B and C. Within the axial recoil approximation the directional emission of these fragments may be used to obtain information on the orientation. Fig. \ref{fig:orientmap}B displays the orientation parameter <cos~$\theta_{\text{exp}}$> versus the kinetic energy of the C$^{2+}$ fragments and the phase $\varphi$, where non-zero values of <cos~$\theta_{\text{exp}}$> indicate a net macroscopic orientation of the molecular ensemble. An oscillatory behavior of <cos~$\theta_{\text{exp}}$> with $\varphi$ is seen throughout the displayed kinetic energy spectrum and <cos~$\theta_{\text{exp}}$> changes its sign at every $\pi$ phase shift between the two colors.
\\
The orientation is effectively controlled by the phase of the two-color laser field. At $\varphi$=0, <cos~$\theta_{\text{exp}}$> is negative, meaning the C$^{2+}$ ions are emitted preferentially downwards, while at $\varphi$=$\pi$, <cos~$\theta_{\text{exp}}$> is positive and more C$^{2+}$ ions are emitted upwards. Note that we recorded similar data for O$^{2+}$ fragments, which show <cos~$\theta_{\text{exp}}$> shifted in $\varphi$ by $\pi$ relative to the C$^{2+}$ data. This is consistent with the charged C and O fragments of the same molecule being emitted in opposite directions within the axial recoil approximation. At phases of $\varphi$=0 and $\varphi$=$\pi$, where <cos~$\theta_{\text{exp}}$> peaks in Fig. \ref{fig:orientmap}B, we have recorded the full time dependence of the orientation parameter. The resulting two curves are shown in Fig. \ref{fig:alor}B. No sign of orientation is found at the half revivals of <cos$^2~\theta_{\text{exp}}$> (seen in Fig. \ref{fig:alor}A). Non-adiabatic field-free orientation manifests itself in Fig. \ref{fig:alor}B in the revivals of <cos~$\theta_{\text{exp}}$> at full rotational periods of CO. In our experiments, we studied the dependence of the degree of orientation on the pump pulse peak intensity and found that orientation decreases rapidly for lower pump intensities.
\\
Our theoretical treatment of the alignment and orientation of CO assumed a quantum mechanical rigid rotor model. In the time-dependent Schr\"{o}dinger equation for the system,
\begin{align*}
i\frac{\partial}{\partial t}\Psi(\theta,t)&=\left[{\bf H}_{0}+E(t)V_{d}(\theta)+E^{2}(t)V_{\rm pol}(\theta)\right.
\nonumber
\\
&\left. + E^{3}(t)V_{\rm hyp}(\theta)\right]\Psi(\theta,t)
\label{TDSE}
\end{align*}
the interaction with the two-color laser field ($E(t)$) took into account the
permanent dipole moment ($V_{d}$), dipole polarizability ($V_{\rm pol}$), and hyperpolarizability ($V_{\rm hyp}$)
contributions as described in Ref. \cite{SakaiJCP2001} with the parameters taken from \cite{Peterson97} and \cite{Pecul05}. The field-free Hamiltonian ${\bf H}_{0}$ and wave function $\Psi(\theta,t)$ satisfy the relation ${\bf H}_{0}\Psi=[B_{0}J(J+1)-D_{\rm e}J^{2}(J+1)^{2}]\Psi$,
with $J$ being the orbital angular momentum quantum number and
$D_{e}=2.79\times10^{-10}$ a.u. being the centrifugal distortion constant.
We expanded $\Psi(\theta,t)$ on spherical harmonics and solved the resulting coupled differential equations in time using a Crank-Nicolson propagation scheme. To achieve convergence of the alignment and orientation parameters, we used angular momenta up to $J$=100 and a time step of 0.2~a.u.. In order to compare with the experiment, the results were thermally averaged over an initial Boltzmann distribution assuming a temperature of 60~K. We used Gaussian pulses of 45~fs FWHM and a smaller intensity (7$\cdot$10$^{13}$W/cm$^2$) than in the experiment to approximately account for volume effects.
\\
Figure \ref{fig:orienttheory}A and \ref{fig:orienttheory}B show the calculated time evolutions of the alignment and orientation parameters
with respect to the pump probe delay, respectively. The revival times for <cos$^2~\theta$> and <cos~$\theta$> agree well with the experimental data. The values of the alignment parameter are significantly lower than in the experimental data, which we attributed earlier to the alignment by the multi-photon probe in the experiment. The modulation of the <cos$^2~\theta$> amplitude at the half and full revivals is significantly higher than in the experiment. However, we found that this amplitude depends critically on the values of the dipole moment and the polarizability, the latter of which is not well known. The theoretical prediction for the orientation trace is in reasonable agreement with the
experimental data. The sign of the orientation is reversed by a change in the phase $\varphi$ of the two-color laser field as also observed in the experiment. Our calculations show that the permanent dipole of CO contributes very little to the orientation of the molecule as has been indicated in Ref. \cite{tehini08}, partly because the dipole moment is very small and partly because the dipole interaction averaged over the fast oscillations of the field vanishes in a many-cycle pulse. It is thus the hyperpolarizability that is responsible for the orientation of CO \cite{tehini08}. Within this model, a phase shift observed in the experiment between the preferential emission of carbon ions in ionization and orientation is in agreement with a positive sign of the hyperpolarizability. Following the statement in Ref. \cite{Peterson97} that the hyperpolarizability and the dipole moment of CO exhibit the same sign, our results are in agreement with the dipole of CO being C$^-$-O$^+$ \cite{Peterson97}.
\\
We have also studied the directional emission of C$^{2+}$ ions in the ionization of CO by the two-color laser field. The recorded C$^{2+}$ ions are found to be preferentially emitted in the direction of the electric field vector at a phase of $\varphi$=0. This finding is consistent with calculations of the ionization of CO \cite{alnaser05_short,Znakovskaya09}. The intensity used for the experimental data shown here is, in fact, above the ionization threshold of CO. Note that we do, however, also observe field-free orientation for lower intensities below the ionization threshold of CO. Significant ionization might lead to depletion of CO molecules in the direction of the electric field vector of the excitation field, corresponding to negative <cos~$\theta_{\text{exp}}$> values at $\varphi$=0. We do expect orientation created by ionization depletion to peak at the time of temporal overlap (defined as time zero in the experimental traces) between the two-color excitation field and the probe. The experimental data clearly shows a temporal shift of 0.125 ps between the temporal overlap of the pump and probe pulses (time zero) and the first peak of the orientation. On this basis we conclude that orientation by ionization depletion is at most minor.
\\
In summary, we have reported on non-adiabatic field-free orientation of a heteronuclear molecule (CO) induced by a strong two-color femtosecond laser field. Rotational revivals of the orientation are found after each full rotational period. At the revivals, an oriented ensemble of heteronuclear molecules in the absence of any external electric field is produced. The approach demonstrated here is applicable for a variety of heteronuclear molecules. These field-free oriented molecules can be used to study angle-differential properties of heteronuclear molecules, such as various ionization and scattering cross-sections and may also enable applications such as high-harmonic generation and electron diffraction studies from oriented molecules. Theoretical calculations indicate that the hyperpolarizability is responsible for the orientation of CO. Although the degree of orientation found in the present work on CO is rather small, higher degrees of orientation are expected for other systems at similar laser parameters \cite{tehini08}. A subject for further investigations is to apply shaped laser fields or tailored excitation pulse sequences in order to increase the degree of orientation by two-color laser fields.
\\
We acknowledge support by the Chemical Sciences, Geosciences and Biosciences Division of the U.S. Department of Energy. I.Z. and M.F.K. are grateful for support by the DFG via the Emmy-Noether program and the Cluster of Excellence "Munich Center for Advanced Photonics".
%
%
\bibliographystyle{apsrev}
\bibliography{Literature}
\end{document}